\documentclass[
a4paper,
showpacs,showkeys,preprintnumbers,aps,prd]{revtex4}

\usepackage{amsmath}
\usepackage{amssymb}
\usepackage{revsymb}
\usepackage{graphicx}
\usepackage{dcolumn}

\begin{document}

\title{Axial charges of the nucleon and $N^*$ resonances}
\author{
Ki-Seok Choi, W. Plessas, and R.\,F. Wagenbrunn}
\affiliation{
Theoretische Physik, Institut f\"ur Physik, Karl-Franzens-Universit\"at,
Universit\"atsplatz 5, A-8010 Graz, Austria}

\begin{abstract}
The axial charges of the nucleon and the well-established $N^*$ resonances are studied
within a consistent framework. For the first time the axial charges of the $N^*$
resonances are produced for the relativistic constituent quark model. The
axial charge of the nucleon is predicted close to experiment, and the ones of
$N^*$(1535) and $N^*$(1650), the only cases where such a comparison is possible, agree
well with results from quantum chromodynamics on the lattice that have recently become
available. The relevance of the magnitudes of the $N^*$ axial charges for the low-energy
behavior of quantum chromodynamics is discussed.    
\end{abstract}

\pacs{11.30.Rd, 12.39.Ki, 14.20.Gk, }
\keywords{Axial charge, Relativistic quark model, Nucleon resonance}

\maketitle
The axial charge $g_A$ of the nucleon ($N$) is an essential quantity in understanding the
electroweak and strong interactions within the Standard Model of elementary particles. 
In the first instance it is directly related to the neutron $\beta$ decay, and
its experimental value can be deduced from the ratio of the axial to the vector coupling
constants $g_A/g_V=1.2695\pm0.0029$~\cite{Amsler:2008zzb}; usually this is done under
the assumption of conserved vector currents (CVC), which implies $g_V=1$. The deviation of
$g_A$ from 1, the axial charge of a point-like particle, can be attributed,
according to the Adler-Weisberger sum rule~\cite{Adler:1965ka,Weisberger:1965hp}, to the
differences between the $\pi^+ N$ and $\pi^- N$ cross sections in pion-nucleon scattering.
Through the Goldberger-Treiman relation, $g_A=f_{\pi}g_{\pi NN}/M_N$, the axial charge
is connected with the $\pi$ decay constant $f_{\pi}$, the $\pi NN$ coupling constant
$g_{\pi NN}$, and the nucleon mass $M_N$~\cite{Goldberger:1958tr}.
Thus the axial charge of the $N$ plays a key role for the spontaneous breaking of chiral
symmetry (SB$\chi$S) of quantum chromodynamics (QCD) in low-energy hadron physics, a
phenomenon that is manifested by the non-vanishing value of the light-flavor chiral
condensate $\left<0|q\bar q|0\right>^{1/3}\approx -235$ MeV. 

There have been a number of theoretical attempts to produce the axial charge of the $N$
ground state with many different methods. We mention only the more novel approaches
via the relativistic constituent quark model
(RCQM)~\cite{Glozman:2001zc,Boffi:2001zb,Merten:2002nz},
by chiral perturbation theory~\cite{Bernard:2007zu}, and within lattice
QCD~\cite{Dolgov:2002zm,Edwards:2005ym,Khan:2006de,Yamazaki:2008py,Lin:2008uz,
Alexandrou:2008rp,Takahashi:2008fy}. In general, the theoretical results come close to
the experimental value of roughly 1.27, with the lattice-QCD predictions scattering
over a range of approximately 1.10$-$1.40, depending on the various actions employed and
a series of technical details entering the calculations by the different groups.

Recently, also the axial charges of the $N^*$ resonances have come into the focus of
interest, as it was suggested that their values should become small or even vanishing
for excited states that could be parity partners in a scenario of chiral-symmetry
restoration higher in the hadron spectra~\cite{Glozman:2007ek,Glozman:2008vg}. As
the $g_A$ values of $N^*$ resonances can hardly be measured experimentally, this
remains a highly theoretical question. However, the problem can be explored by
ab-initio calculations of QCD on the lattice. Corresponding first results have become
available lately for just two of the $N^*$ resonances, namely, $N^*$(1535) and
$N^*$(1650)~\cite{Takahashi:2008fy}. Both of them have total angular momentum
(intrinsic spin) $J=\frac{1}{2}$ and parity $P=-1$. Unfortunately, there is not yet
any lattice-QCD result for positive-parity states, and the above issue relating to
parity-doubling remains unresolved on this basis.

It is thus most interesting to get insight into the $N$ and $N^*$ axial charges from
other approaches. Especially by the RCQM we can investigate the problem in a
comprehensive manner,
as all the ground and resonant states are readily accessible. Here we report
theoretical predictions of $g_A$ for positive- as well as negative-parity $N^*$ resonances
up to $J=\frac{5}{2}$. The calculations are performed employing a RCQM with the quark-quark hyperfine interaction
deduced from Goldstone-boson-exchange (GBE) dynamics. In particular, we use both
existing versions of GBE RCQMs, the one with pseudoscalar (ps) spin-spin forces
only~\cite{Glozman:1997ag}
and the extended GBE (EGBE) RCQM with all the central, spin-spin, tensor, and
spin-orbit force components included~\cite{Glantschnig:2004mu}. For the sake of
comparison with another type of hyperfine interaction we employ also the RCQM with
one-gluon-exchange (OGE) dynamics~\cite{Theussl:2000sj}.

The calculations are performed in the framework of Poincar\`e-invariant quantum mechanics.
In order to keep the numerical computations manageable, we have to restrict the axial 
current operator to the so-called spectator model (SM). It means that the weak-interaction
gauge boson couples only to one of the constituent quarks in the baryon. This approximation
has turned out to be very reasonable already in a previous study of the axial and induced
pseudoscalar form factors of the nucleon~\cite{Glozman:2001zc}, where the SM was employed
specifically in the point form (PF) of relativistic quantum mechanics~\cite{Melde:2004qu}.
It has also been used in studies of the electromagnetic structure of the $N$,
reproducing both the proton and neutron form factors in close agreement with the
experimental data~\cite{Boffi:2001zb,Wagenbrunn:2000es,Berger:2004yi,Melde:2007zz}.

The axial charge is defined through the value of the axial form factor $G_A(Q^2)$ at
$Q^2=0$, where $Q^2=-q^2$ is the four-momentum transfer. The axial form factor
$G_A(Q^2)$ can be deduced from the invariant matrix element of the axial current
operator $\hat A^\mu_a (Q^2)$, with flavor index $a$, sandwiched between
the eigenstates of $N$ or $N^*$. We denote the latter generally by
$\left|P,J,\Sigma\right>$, i.e. as
eigenstates of the four-momentum operator $\hat P^\mu$, the intrinsic-spin operator
$\hat J$ and its $z$-projection $\hat \Sigma$. Since $\hat P^\mu$ and the invariant mass
operator $\hat M$ commute, these eigenstates can be obtained by solving the eigenvalue
equation of $\hat M$
\begin{equation}
\hat M \left|P,J,\Sigma\right>=M \left|P,J,\Sigma\right> \, ,
\end{equation}
where $M$ is the mass of $N$ or $N^*$. For the various
$J=\frac{1}{2}, \frac{3}{2}, \frac{5}{2}$ states considered here, the axial charges
$g_A$ are thus computed from the matrix elements of the axial current operator
$\hat A^\mu_a$ for zero momentum transfer
\begin{eqnarray}
\left<P,\frac{1}{2},\Sigma'\left|{\hat A}^{\mu}_a\right|P,\frac{1}{2},\Sigma\right>&=&
\bar U(P,\Sigma')g_A \gamma^{\mu}\gamma_5 \frac{\tau_a}{2} U(P,\Sigma) \, ,\nonumber \\
\left<P,\frac{3}{2},\Sigma'\left|{\hat A}^{\mu}_a\right|P,\frac{3}{2},\Sigma\right>&=&
\bar U^\nu(P,\Sigma')g_A \gamma^{\mu}\gamma_5 \frac{\tau_a}{2} U_\nu(P,\Sigma) \, ,
\nonumber \\
\left<P,\frac{5}{2},\Sigma'\left|{\hat A}^{\mu}_a\right|P,\frac{5}{2},\Sigma\right>&=&
\bar U^{\nu\lambda}(P,\Sigma')g_A \gamma^{\mu}\gamma_5 \frac{\tau_a}{2}
U_{\nu\lambda}(P,\Sigma) \, .\nonumber \\
\end{eqnarray}
Here $U(P,\Sigma)$ are the Dirac spinors for spin-$\frac{1}{2}$ and 
$U^\nu(P,\Sigma)$ as well as $U^{\nu\lambda}(P,\Sigma)$ the Rarita-Schwinger spinors
for spin-$\frac{3}{2}$ and spin-$\frac{5}{2}$ particles, respectively. $\gamma^\mu$ and
$\gamma_5$ are the usual Dirac matrices and $\tau_a$ the isospin matrix with Cartesian
index $a$.

The matrix elements of $\hat A^\mu_a$ for any $N$ or $N^*$ read
\begin{eqnarray}
&&\left<P,J,\Sigma'\right|{\hat A^\mu_a (Q^2=0)}\left|P,J,\Sigma\right>= \nonumber \\
&&2M\sum_{\sigma_i\sigma'_i}{\int{
d^3{\vec k}_1 d^3{\vec k}_2 d^3{\vec k}_3}} 
\frac{\delta^3(\vec k_1+\vec k_2+\vec k_3)}{2\omega_1 2\omega_2 2\omega_3}
\Psi^\star_{PJ\Sigma'}\left({\vec k}_1,{\vec k}_2,{\vec k}_3;
\sigma'_1,\sigma'_2,\sigma'_3\right) \nonumber \\
&&
\times\left<k_1,k_2,k_3;\sigma'_1,\sigma'_2,\sigma'_3\right|\hat{A}^{\mu}_a
\left|k_1,k_2,k_3;\sigma_1,\sigma_2,\sigma_3\right>
\Psi_{PJ\Sigma}\left({\vec k}_1,{\vec k}_2,{\vec k}_3;
\sigma_1,\sigma_2,\sigma_3\right)  \, .
\label{transampl}
\end{eqnarray}
The $\Psi$'s are the rest-frame wave functions of the $N$ or $N^*$ with corresponding
mass $M$ and total angular momentum $J$ with $z$-projections $\Sigma$ and $\Sigma'$.
Here they are represented as functions of the individual quark
three-momenta $\vec k_i$, which sum up to $\vec P=\vec k_1+\vec k_2+\vec k_3=0$;
$\omega_i=\sqrt{m^2_i+\vec k^2_i}$ is the energy of quark $i$ with mass $m_i$, and the
individual-quark spin orientations are denoted by $\sigma_i$.

The SM means that the matrix element of the axial current operator
$\hat{A}^{\mu}_a$ between (free) three-particle states
$\left|k_1,k_2,k_3;\sigma_1,\sigma_2,\sigma_3\right>$  is assumed in the form
%
\begin{equation}
\left<k_1,k_2,k_3;\sigma'_1,\sigma'_2,\sigma'_3\right|
{\hat A}^{\mu}_a
\left|k_1,k_2,k_3;\sigma_1,\sigma_2,\sigma_3\right>=
3\left<k_1,\sigma'_1\right|\hat{A}^{\mu}_{a,{\rm SM}}
\left|k_1,\sigma_1\right>2\omega_2 2\omega_3
\delta_{\sigma_{2}\sigma'_{2}}\delta_{\sigma_{3}\sigma'_{3}}
\label{eq:axcurr1}\, .
\end{equation}
%
For point-like quarks this matrix element involves the axial current operator
of the active quark 1 (with quarks 2 and 3 being the spectators) in the form
%
\begin{equation}
\left<k_1,\sigma'_1\right|\hat{A}^{\mu}_{a,{\rm SM}}
\left|k_1,\sigma_1\right>=
{\bar u}\left(k_1,\sigma'_1\right)g_A^q \gamma^\mu
\gamma_5 \frac{{\tau}_a}{2} u\left(k_1,\sigma_1\right) \, ,
\label{eq:axcurr2}
\end{equation}
%
where $u\left(k_1,\sigma_1\right)$ is the quark spinor and $g_A^q=1$ the quark axial charge.
A pseudovector-type current analogous to the one in Eq.~(\ref{eq:axcurr2}) was recently
also used in the calculation of $g_{\pi NN}$ and the strong $\pi NN$ vertex form factor
in ref.~\cite{Melde:2008dg}.

If we are interested only in the axial charges $g_A$, the expression
(\ref{eq:axcurr2}) specifies to $\mu=i=1, 2, 3$ and can further be evaluated to give 
%
\begin{equation}
{\bar u}\left(k_1,\sigma'_1\right) \gamma^i\gamma_5 \frac{{\tau}_a}{2} u\left( k_1,\sigma_1\right)=
2\omega_1\chi^*_{\frac{1}{2},\sigma'_1}
\Biggl\{\left[1-\frac{2}{3}\left(1-\kappa\right) \right]\sigma^i
+\sqrt{\frac{5}{3}}\frac{\kappa^2}{1+\kappa}
\left[\,\left[\vec{v}_{1}\otimes\vec{v}_{1}\right]_2\otimes\vec{\sigma}\right]_1^i\Biggl\}
\frac{{\tau}_a}{2} \chi_{\frac{1}{2},\sigma_1}
\label{eq:ga}
\end{equation}
%
where $\kappa=1/\sqrt{1+v_1^2}$ and $\vec v_1=\vec k_1/m_1$.
Herein $\sigma^i$ is the $i$-th component of the usual Pauli matrix
$\vec \sigma$ and $v_1$ the magnitude of the three-velocity $\vec v_1$. The symbol
$\left[.\otimes .\right]_k^i$ denotes the $i$-th component of a tensor product
$\left[.\otimes .\right]_k$ of rank $k$. We note that a similar formula was already
published before by Dannbom et al.~\cite{Dannbom:1996sh}, however, restricted to the
case of total orbital angular momentum $L=0$. Our expression holds for any $L$, thus
allowing to calculate $g_A$ for the most general wave function of a baryon specified
by $J^P$.

In Table~\ref{EGBE} we give the predictions of the EGBE RCQM for the $N$ ground state and
the first two $N^*$ excitations of $J=\frac{1}{2}$ with positive as well as negative parity
$P$. It is seen that the result for $g_A$ of the $N$ comes close to the experimental
value and is also congruent with the lattice-QCD results. The latter is also true with
respect to the other cases, where lattice-QCD results are already available, the 
$J^P=\frac{1}{2}^-$ resonances $N^*$(1535) and $N^*$(1650). The simple
$SU(6)\times O(3)$ nonrelativistic quark model used by Glozman and Nefediev cannot
reproduce the $g_A$ of the $N$ and it yields exactly the same results for $N$ and
$N^*$(1440). The corresponding axial charge of $N^*$(1535) is non-zero but negative,
while the results
for $N^*$(1710) and $N^*$(1650) are similar to the ones of the EGBE RCQM. In the last
column of Table~\ref{EGBE} we also quote the results obtained in the nonrelativistic
limit of the axial current operator of Eq.~(\ref{eq:axcurr2}). By comparing with the
figures in the first column, one can see that relativistic effects related to the
current operator are considerable in all instances.

\renewcommand{\arraystretch}{1.4}
\begin{table}[h]
\centering
\caption{Predictions for axial charges $g_A$ of the EGBE RCQM in comparison to available
lattice QCD results~\cite{Dolgov:2002zm,Edwards:2005ym,Khan:2006de,Yamazaki:2008py,
Lin:2008uz,Alexandrou:2008rp,Takahashi:2008fy},
the values calculated by Glozman and Nefediev (GN) within the
$SU(6)\times O(3)$ nonrelativistic quark model, and the nonrelativistic (NR) limit from the
EGBE RCQM.} 
\begin{tabular}{lcp{2mm}cp{2mm}cp{2mm}rp{2mm}r}
\hline\hline \\[-3mm]
State & $J^P$ && EGBE && Lattice QCD && GN && NR\\
\hline \\[-2mm]
    $N$(939) & $\frac{1}{2}^{+}$ & & 1.15 && 1.10$\sim$1.40 && 1.66 && 1.65 \\ 
    $N^*$(1440) & $\frac{1}{2}^{+}$ & & 1.16 && -- && 1.66 && 1.61 \\
    $N^*$(1535) & $\frac{1}{2}^{-}$ & & 0.02 && $\sim$0.00 && -0.11 && -0.20 \\
    $N^*$(1710) & $\frac{1}{2}^{+}$ & & 0.35 && -- && 0.33 && 0.42 \\
    $N^*$(1650) & $\frac{1}{2}^{-}$ & & 0.51 && $\sim$0.55 && 0.55 && 0.64 \\[1mm]
\hline\hline
\end{tabular}
\label{EGBE}
\end{table}

In Tables~\ref{POS} and~\ref{NEG} we present the relativistic predictions of $g_A$ for
the $N$ ground state and all positive- as well as negative-parity $N^*$ excitations with
masses below $\sim$1.9 GeV for the three types of RCQMs considered here. We regard
the EGBE result as the most advanced one, as this particular RCQM includes all force
components in the hyperfine interaction and presumably produces the most realistic
$N$ and $N^*$ wave functions. Still, the psGBE RCQM, which relies only
on a spin-spin hyperfine interaction, performs similarly
well for all positive-parity resonances and for most of the negative-parity resonances;
only for the $J^P=\frac{3}{2}^-$ states $N^*$(1520) and $N^*$(1700) there occur
differences, which have evidently to be attributed to tensor and/or spin-orbit forces.
Except for these two cases there are also no big differences to the results with the
OGE RCQM, even though the theoretical resonance masses show sometimes considerable
differences~\cite{Theussl:2000sj,Plessas:2003}.
It will be very interesting to confront these predictions by the RCQMs with results
by other approaches and in particular with further data from lattice QCD.

\begin{table}[h]
\centering
\caption{Mass eigenvalues and axial charges $g_A$ of the $N$ ground state and the
positive-parity $N^*$ resonances as predicted by the EGBE, the psGBE, and the OGE RCQMs.}
\begin{tabular}{lcp{3mm}crp{3mm}crp{3mm}cr}
\hline\hline \\[-3mm]
\multicolumn{2}{c}{} && \multicolumn{2}{c}{EGBE} && \multicolumn{2}{c}{psGBE} && \multicolumn{2}{c}{OGE}\tabularnewline
\hline 
State & $J^{p}$ && Mass & $g_{A}$ && Mass & $g_{A}$ && Mass & $g_{A}$\tabularnewline
\hline\\[-2mm]  
$N$(939) & $\frac{1}{2}^{+}$ && 939 & 1.15 && 939& 1.15 && 939 & 1.11\tabularnewline

$N^*$(1440) & $\frac{1}{2}^{+}$ && 1464 & 1.16 && 1459 & 1.13 && 1578 & 1.10\tabularnewline

$N^*$(1710) & $\frac{1}{2}^{+}$ && 1757 & 0.35 && 1776 & 0.37 && 1860 & 0.32\tabularnewline

$N^*$(1720) & $\frac{3}{2}^{+}$ && 1746 & 0.35 && 1728 & 0.34 && 1858 & 0.25\tabularnewline

$N^*$(1680) & $\frac{5}{2}^{+}$ && 1689 & 0.89 && 1728& 0.83 && 1858 & 0.70\tabularnewline
\hline
\hline
\end{tabular}
\label{POS}
\end{table}

\begin{table}[h]
\centering
\caption{Same as Table~\ref{POS} but for the negative-parity $N^*$ resonances.}
\begin{tabular}{lcp{3mm}crp{3mm}crp{3mm}cr}
\hline\hline \\[-3mm] 
\multicolumn{2}{c}{} && \multicolumn{2}{c}{EGBE} && \multicolumn{2}{c}{psGBE} && \multicolumn{2}{c}{OGE}\tabularnewline
\hline
State & $J^{p}$ && Mass & $g_{A}$ && Mass & $g_{A}$ && Mass & $g_{A}$\tabularnewline
\hline\\[-2mm] 
$N^*$(1535) & $\frac{1}{2}^{-}$ && 1498 & 0.02 && 1519 & 0.09 && 1520 & 0.13\tabularnewline

$N^*$(1650) & $\frac{1}{2}^{-}$ && 1581 & 0.51 && 1647 & 0.46 && 1690 & 0.44\tabularnewline

$N^*$(1520) & $\frac{3}{2}^{-}$ && 1524 & -0.64 && 1519 & -0.21 && 1520 & -0.15\tabularnewline

$N^*$(1700) & $\frac{3}{2}^{-}$ && 1608 & -0.10 && 1647 & -0.50 && 1690 & -0.47\tabularnewline

$N^*$(1675) & $\frac{5}{2}^{-}$ && 1676 & 0.84 && 1647 & 0.83 && 1690 & 0.80\tabularnewline
\hline
\hline
\end{tabular}
\label{NEG}
\end{table}

While the $g_A$ of the nucleon as predicted by the RCQM falls well into the interval of
values reported from lattice QCD, it is nevertheless smaller than the experimental
value of $\sim$1.27, supposed under the conjecture of CVC. It is not yet clear what is
the reason for this undershooting of the nucleon's $g_A$ by the RCQM. For example,
it could well be that CVC is violated
up to 10\% or the assumption of $g_A^q=1$ for the constituent quarks is not
justified. Further investigations are necessary to clarify these questions.

It is particularly satisfying to find the RCQM predictions for the axial charges
not only of the nucleon $N$ but also of the
$N^*$(1535) and $N^*$(1650) resonances in agreement with the lattice-QCD results. We
may thus be confident that at least in the limit of
zero momentum-transfer processes the mass
eigenstates of the $N$ and the above $N^*$ excitations, especially as produced with
the EGBE RCQM, are quite reasonable. It should be recalled that in this particular model
the mutual interaction between constituent quarks is furnished by a linear confinement,
whose strength is consistent with the string tension of QCD as well as the slopes of Regge
trajectories~\cite{Glozman:1997ag}, and by (all components of) a hyperfine interaction
derived from GBE dynamics. The latter is supposed to model
the SB$\chi$S property of low-energy QCD. This type of hyperfine interaction, which
also introduces an explicit flavor dependence, has been remarkably successful in
describing a number of phenomena in low-energy baryon physics. Most prominently,
it produces the correct level orderings of the positive- and negative-parity $N^*$
resonances and simultaneously the ones in the other hyperon spectra, notably the
$\Lambda$ spectrum~\cite{Melde:2008yr}.
The RCQM with GBE dynamics does not have any mechanism for
chiral-symmetry restoration built in. As such it cannot be expected to produce
parity doublets due to this reason. Nevertheless the GBE RCQM describes the pattern of
$N^*$ resonance masses in good agreement with the experimental data (mostly within
the experimental error bars or at most exceeding them by 4\%). This is due to the refined
interplay of different force components in the effective interaction between constituent
quarks. In view of these findings
it will be most interesting if the present results for $N^*$ axial charges derived
within the RCQM will in the future be confirmed by lattice-QCD calculations.

\begin{acknowledgments}
This work was supported by the Austrian Science Fund (through the Doctoral
Program on {\it Hadrons in Vacuum, Nuclei, and Stars}; FWF DK W1203-N08). The authors are
grateful to L. Glozman for suggestive ideas and a number of clarifying discussions.

\end{acknowledgments}


\addcontentsline{toc}{chapter}{Bibliography}
\bibliographystyle{prsty}

\begin{thebibliography}{10}

\bibitem{Amsler:2008zzb}
  C.~Amsler {\it et al.}  [Particle Data Group],
  Phys.\ Lett.\  B {\bf 667}, 1 (2008).
  
  \bibitem{Adler:1965ka}
  S.~L.~Adler,
  Phys.\ Rev.\ Lett.\  {\bf 14}, 1051 (1965).

\bibitem{Weisberger:1965hp}
  W.~I.~Weisberger,
  Phys.\ Rev.\ Lett.\  {\bf 14}, 1047 (1965).
  
  \bibitem{Goldberger:1958tr}
  M.~L.~Goldberger and S.~B.~Treiman,
  Phys.\ Rev.\  {\bf 110}, 1178 (1958).

\bibitem{Glozman:2001zc}
  L.~Y.~Glozman, M.~Radici, R.~F.~Wagenbrunn, S.~Boffi, W.~Klink and W.~Plessas,
  Phys.\ Lett.\  B {\bf 516}, 183 (2001).
  
\bibitem{Boffi:2001zb}
  S.~Boffi, L.~Y.~Glozman, W.~Klink, W.~Plessas, M.~Radici and R.~F.~Wagenbrunn,
  Eur.\ Phys.\ J.\  A {\bf 14}, 17 (2002).
  
\bibitem{Merten:2002nz}
  D.~Merten, U.~Loring, K.~Kretzschmar, B.~Metsch and H.~R.~Petry,
  Eur.\ Phys.\ J.\  A {\bf 14}, 477 (2002).  

\bibitem{Bernard:2007zu}
  V.~Bernard,
  Prog.\ Part.\ Nucl.\ Phys.\  {\bf 60}, 82 (2008).

\bibitem{Dolgov:2002zm}
  D.~Dolgov {\it et al.}  [LHPC collaboration and TXL Collaboration],
  Phys.\ Rev.\  D {\bf 66}, 034506 (2002).

\bibitem{Edwards:2005ym}
  R.~G.~Edwards {\it et al.}  [LHPC Collaboration],
  Phys.\ Rev.\ Lett.\  {\bf 96}, 052001 (2006).

\bibitem{Khan:2006de}
  A.~A.~Khan {\it et al.},
  Phys.\ Rev.\  D {\bf 74}, 094508 (2006).

\bibitem{Yamazaki:2008py}
  T.~Yamazaki {\it et al.}  [RBC+UKQCD Collaboration],
  Phys.\ Rev.\ Lett.\  {\bf 100}, 171602 (2008).

\bibitem{Lin:2008uz}
  H.~W.~Lin, T.~Blum, S.~Ohta, S.~Sasaki and T.~Yamazaki,
  Phys.\ Rev.\  D {\bf 78}, 014505 (2008).

\bibitem{Alexandrou:2008rp}
  C.~Alexandrou {\it et al.}  PoS(LATTICE 2008), 139 (2008);
  arXiv:0811.0724 [hep-lat].

\bibitem{Takahashi:2008fy}
  T.~T.~Takahashi and T.~Kunihiro,
  Phys.\ Rev.\  D {\bf 78}, 011503 (2008).

\bibitem{Glozman:2007ek}
  L.~Y.~Glozman,
  Phys.\ Rept.\  {\bf 444}, 1 (2007).

\bibitem{Glozman:2008vg}
  L.~Y.~Glozman and A.~V.~Nefediev,
  Nucl.\ Phys.\  A {\bf 807}, 38 (2008).
 
\bibitem{Glozman:1997ag}
  L.~Y.~Glozman, W.~Plessas, K.~Varga and R.~F.~Wagenbrunn,
  Phys.\ Rev.\  D {\bf 58}, 094030 (1998).

\bibitem{Glantschnig:2004mu}
  K.~Glantschnig, R.~Kainhofer, W.~Plessas, B.~Sengl and R.~F.~Wagenbrunn,
  Eur.\ Phys.\ J.\  A {\bf 23}, 507 (2005).

\bibitem{Theussl:2000sj}
  L.~Theussl, R.~F.~Wagenbrunn, B.~Desplanques and W.~Plessas,
  Eur.\ Phys.\ J.\  A {\bf 12}, 91 (2001).
  
\bibitem{Melde:2004qu}
  T.~Melde, L.~Canton, W.~Plessas and R.~F.~Wagenbrunn,
  Eur.\ Phys.\ J.\  A {\bf 25}, 97 (2005).

\bibitem{Wagenbrunn:2000es}
  R.~F.~Wagenbrunn, S.~Boffi, W.~Klink, W.~Plessas and M.~Radici,
  Phys.\ Lett.\  B {\bf 511}, 33 (2001).

\bibitem{Berger:2004yi}
  K.~Berger, R.~F.~Wagenbrunn and W.~Plessas,
  Phys.\ Rev.\  D {\bf 70}, 094027 (2004).

\bibitem{Melde:2007zz}
  T.~Melde, K.~Berger, L.~Canton, W.~Plessas and R.~F.~Wagenbrunn,
  Phys.\ Rev.\  D {\bf 76}, 074020 (2007).

\bibitem{Melde:2008dg}
  T.~Melde, L.~Canton and W.~Plessas,
  Phys.\ Rev.\ Lett.\  {\bf 102}, 132002 (2009).

\bibitem{Dannbom:1996sh}
  K.~Dannbom, L.~Y.~Glozman, C.~Helminen and D.~O.~Riska,
  Nucl.\ Phys.\  A {\bf 616}, 555 (1997).
  
\bibitem{Plessas:2003}
  W.~Plessas, Few-Body Syst. Suppl. {\bf 15}, 139 (2003). 
  
\bibitem{Melde:2008yr}
  T.~Melde, W.~Plessas and B.~Sengl,
  Phys.\ Rev.\  D {\bf 77}, 114002 (2008).
  
    
\end{thebibliography}

\end{document}